\documentclass[aps,prm,twocolumn,floatfix,10pt]{revtex4}
\usepackage{graphicx}
\usepackage{amsmath}
\usepackage{color}
\usepackage{wrapfig}
\usepackage{latexsym}
\begin{document}

%\title{Solvation thermodynamics of commodity polymer blends: An analysis based on the Kirkwood-Buff theory of mixtures}
%\title{Kirkwood-Buff analysis of commodity polymer blends}
\title{Correlating thermodynamics, morphology, mechanics and thermal transport in PMMA-PLA blends}

\author{Debashish Mukherji}
\email[]{debashish.mukherji@ubc.ca}
\affiliation{Quantum Matter Institute, University of British Columbia, Vancouver BC V6T 1Z4, Canada}
\author{Tiago Espinosa de Oliveira}
\affiliation{Departamento de Farmacoci\^encias, Universidade Federal de Ci\^encias da Sa\'ude de Porto Alegre, Porto Alegre 90050-170, Brazil}
\author{C\'eline Ruscher}
\affiliation{Department of Mechanical Engineering, University of British Columbia, Vancouver BC V6T 1Z4, Canada}
\author{J\"org Rottler}
\affiliation{Quantum Matter Institute, University of British Columbia, Vancouver BC V6T 1Z4, Canada}
\affiliation{Department of Physics and Astronomy, University of British Columbia, Vancouver BC V6T 1Z1, Canada}

\date{\today}

\begin{abstract}
Thermodynamics controls structure, function, stability and morphology of polymer blends. 
However, obtaining the precise information about their mixing thermodynamics is a challenging task, 
especially when dealing with complex macromolecules.
This is partially because of a delicate balance between the local concentration/composition fluctuations 
and the monomer level (multi-body) interactions. In this context, the Kirkwood-Buff (KB) theory serves 
as a useful tool that connects the local pairwise fluid structure to the mixing thermodynamics. 
Using larger scale molecular dynamics simulations, within the framework of KB theory, 
we investigate a set of technologically relevant poly(methyl methacrylate)-poly(lactic acid) (PMMA-PLA) blends with the aim to elucidate
the underlying microscopic picture of their phase behavior. %The specific choice of PMMA-PLA blends is
%motivated by a recent experiment that reported an extraordinary mechanical response, i.e.,
%with the increasing concentration of PLA a system becomes more ductile.
Consistent with these experiments, we emphasize the importance of properly accounting for the entropic contribution, 
to the mixing Gibbs free energy change $\Delta {\mathcal G}_{\rm mix}$, that controls the phase morphology. 
We further show how the relative microscopic interaction details and the molecular level structures between different mixing 
species can control the non-linear mechanics and ductility. 
%We argue based on the monomer structures and polymer
%flexibilities that provide a possible explanation for the mechanical response in these materials.
As a direct consequence, we provide a correlation that links thermodynamics, phase behavior, mechanics, and 
thus also thermal transport in polymer blends. Therefore, this study provides 
a guiding principle for the design of light weight functional materials with extraordinary physical properties.
\end{abstract}

\maketitle

\section{Introduction}
\label{sec:intro}

Polymers are an important class of soft materials that are extensively used for the advanced applications \cite{Kroeger04PR,Mueller20PPS,Mukherji20AR},
this reaches from common household items to the advanced technologies for their use in mechanical \cite{stevens01mac,mukherji08pre,white10anrev,palmese14jmat}, 
thermal \cite{pipe14nm,xie16mac,mukherji19mac} and optoelectronics devices \cite{kim05apl,cola16aami}. These systems consist of, but are not
limited to, linear polymers, copolymers, polymers with different architectures and/or polymer blends. 
In this context, the latter is of particular interest because a polymer blend provides an additional 
flexibility to design functional materials with tunable properties \cite{pipe14nm,mukherji19mac,hou21maclet}. Here, however, the microscopic 
phase behavior is important for the usefulness of these materials for their desired applications \cite{paul03,yeo20}. 
For example, the micro-phase separation in these systems severely impacts the materials performance 
by creating the weak spots that act as the nucleating sites for the 
crack propagation and mechanical instabilities in these materials \cite{paul03}. Therefore, it is important to achieve a 
compatible chemical composition that is not only simple from the processing point of view, but also 
provides a reasonably stable homogeneous blend. Here, the standard commodity polymers, such as poly(methyl methacrylate)
(PMMA), poly($N-$acryloyl piperidine) (PAP), poly(lactic acid) (PLA), poly(acrylic acid) (PAA), and poly(acrylamide) (PAM), 
have been shown to provide extraordinary phase, mechanical, and thermal behavior 
\cite{pipe14nm,xie16mac,mukherji19mac,le06,sam15,gar18,hou21maclet,paul03,yeo20}.

The miscibility of polymer blends at different mixing ratios is dictated by the change in Gibbs free energy
of mixing $\Delta {\mathcal G}_{\rm mix} = \Delta H_{\rm mix} - T\Delta S_{\rm mix}$, where 
$\Delta H_{\rm mix}$ and $T\Delta S_{\rm mix}$ are the enthalpic and the entropic contributions, respectively.
A positive value of $\Delta {\mathcal G}_{\rm mix}$ does not favor miscibility, while miscibility is favored 
for $\Delta {\mathcal G}_{\rm mix} < 0$.
Within the simple Flory-Huggins (FH) like representation \cite{doibook,genbook}, one can estimate the entropy change $\Delta S_{\rm mix}$ as,
\begin{equation}
\label{eq:entropy}
\Delta S_{\rm mix} = -k_{\rm B}\left[\frac{x_{\rm i}}{N_{\rm i}} \ln{x_{\rm i}} + \frac{1-x_{\rm i}}{N_{\rm j}}\ln\left({1-x_{\rm i}}\right) \right].
\end{equation}
Here, $k_{\rm B}$ is the Boltzmann constant, ${x_{\rm i}}$ is mole fraction of species i, and $N_{\rm i}$ is the degree of polymerization of species i.
While $\Delta S_{\rm mix}$ scales as $1/N$, it is always positive and thus favors miscibility. On the contrary, when 
$\Delta H_{\rm mix} = \chi {x_{\rm i}} \left({1-x_{\rm i}}\right)  >> 0$ (or $\chi >> 0$),
a system can phase separate, with $\chi$ being the FH interaction parameter.

Following the discussion in the preceding paragraph, it becomes apparent that a delicate balance between $\Delta H_{\rm mix}$ and $T\Delta S_{\rm mix}$
plays a key role in dictating the phase behavior of polymer blends. Therefore, it is crucial to obtain these individual 
contributions to $\Delta {\mathcal G}_{\rm mix}$, which is a nontrivial task in molecular simulations. 
In this context, most computational studies usually deal with enthalpy ($\chi-$based arguments),
while ignoring $T\Delta S_{\rm mix}$ contributions. %to $\Delta {\mathcal G}_{\rm mix}$.
Here, we note in passing that most molecular simulations of polymers commonly deal with the coarse-grained (CG) potentials
that are by construction free energies (not internal energies) and thus contain both the entropic and the enthalpic contributions. Therefore,
special attention should be devoted when discussing results using different CG models.

Studying the thermodynamic properties of polymer mixtures is a nontrivial task within molecular simulation protocol.
Here, the fluctuation theory of Kirkwood-Buff (KB) serves as an important method, 
where local fluctuations and monomer-level interactions play a key role in dictating the mixing thermodynamics.
While the KB theory has been extensively used in studying various solvent mixtures \cite{kbi,kbibook,robin16jcp,doros18fluid,robin21arx}, 
polymer solutions \cite{smith07,pier08,mukherji13mac}, and/or ionic systems \cite{kumari21pccp}, 
their application for the systems with long chain blends are rather limited, if existing at all.
This is in particular because in polymers, chain connectivity introduces additional constraints 
that significantly modifies the solvation structure, which is otherwise spherically symmetric,
and thus can significantly impact their solvation thermodynamics.

In this work, we investigate phase behavior and its links to the nonlinear mechanics and the 
thermal transport in a set of PMMA-PLA blends at different PLA mixing ratios $x_{\rm PLA}$.
The specific system choice is motivated by the recent experimental study \cite{hou21maclet}, 
where it has been shown that the PMMA-PLA blends exhibit extraordinary mechanical response 
and ultra-high toughness. We individually identify the entropic and the enthalpic contributions 
to highlight the thermodynamic origin of their phase behavior. We use the phase behavior to establish a delicate link
with the mechanics of polymer blends and ultimately to the thermal transport.

The remainder of the paper is as follows: In Sec. \ref{sec:meth} we sketch brief description of the KB theory, model and method related details and
the materials used in our study. Results are presented in Sec. \ref{sec:res} and finally we draw our conclusions in Sec. \ref{sec:conc}.

\section{Method, model and materials}
\label{sec:meth}

\subsection{Kirkwood-Buff theory of solution}

KB theory is a well known method to investigate the thermodynamics properties of complex mixtures. Here
we only sketch the key ingredients relevant for our work. When two different species i and j are blended in at different $x_{\rm j}$, 
the fluctuation theory of Kirkwood and Buff connects fluctuations in the grand-canonical ensemble to the pair distribution functions 
${\rm g}_{\rm ij}(r)$ via ``so called" Kirkwood-Buff integral (KBI) \cite{kbi,kbibook},
\begin{equation}
\label{eq:kbi}
        G_{\rm ij} = 4 \pi \int_0^{\infty} \left[ {\rm g}_{\rm ij}(r) - 1\right]r^2 dr.
\end{equation}
Here $G_{\rm ij}$ is known as the excess coordination and in polymer physics $-G_{\rm ij}/2$ gives an 
estimate of the monomer excluded volume ${\mathcal V}_{\rm ij}$. 

Following Eq.~\ref{eq:kbi}, the second derivative of $\Delta {\mathcal G}_{\rm mix}$ can be written in terms of $G_{\rm ij}$ 
as \cite{doros18fluid},
\begin{equation}
\label{eq:gibbs}
	\frac {\partial^2 \Delta {\mathcal G}_{\rm mix}}{\partial x_{\rm j}^2} = \frac {k_{\rm B}T \left(\rho_{\rm i} + \rho_{\rm j}\right)} 
	{x_{\rm j} (1- x_{\rm j}) \left(\rho_{\rm j} + \rho_{\rm i} + \rho_{\rm j} \rho_{\rm i} \eta \right)},
\end{equation}
with $\rho_{\rm i}$ being the monomer number density of species i. 
The preferential solvation parameter $\eta = G_{\rm ii} + G_{\rm jj} - 2 G_{\rm ij}$ gives further information about the relative (local) 
coordination between different mixture components. For example, immiscibility is preferred when $G_{\rm ii} + G_{\rm jj} \gg 2 G_{\rm ij}$,
weakly immiscible to miscible when $G_{\rm ii} + G_{\rm jj} \geq 2 G_{\rm ij}$
the system is miscible for $\eta \le 0$. Note that $G$ values depend on the volume of the molecules.

Eq.~\ref{eq:gibbs} satisfies the boundary condition $\Delta {\mathcal G}_{\rm mix}(x_{\rm j} = 0.0) = 0$ and $\Delta {\mathcal G}_{\rm mix}(x_{\rm j} = 1.0) = 0$.
From $\Delta {\mathcal G}_{\rm mix}$, one can estimate $T \Delta S_{\rm mix} = \Delta H_{\rm mix} - \Delta {\mathcal G}_{\rm mix}$ with,
\begin{equation}
\label{eq:enthalpy}
       \Delta H_{\rm mix} = H_{\rm mixture} - x_{\rm j} H_{\rm j}^{\rm pure} - \left(1 - x_{\rm j}\right) H_{\rm i}^{\rm pure}.
\end{equation}
Here, the enthalpy per monomer is calculated using $H = U +pV$, $U$ is the internal energy including the kinetic energy, $p$
is the externally imposed pressure and $V$ is the system volume.

\subsection{All atom simulations}

\begin{figure}[h!]
	\includegraphics[width=0.25\textwidth,angle=0]{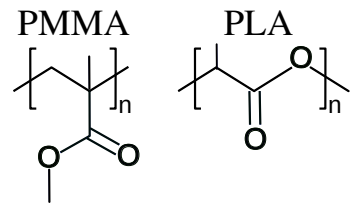}
        \caption{Schematics showing the chemical structures of the monomer units investigated in this work: namely poly(methyl methacrylate) (PMMA)
	and poly(lactic acid) (PLA).
\label{fig:schem}}
\end{figure}

For the all-atom simulations, we have chosen two different polymers: namely PMMA and PLA.
The specific chemical structures are shown in Fig. \ref{fig:schem}. 
The explicit atom OPLS force field is used to model PLA\cite{OPLS}, while we have used a set of modified 
OPLS parameters for PMMA \cite{Mukherji17NC}. Earlier it has been shown that the PMMA force field parameters 
reasonably capture the properties of the solid, solvent-free PMMA, 
such as the specific volume, the macroscopic elastic moduli, and the thermal transport coefficient \cite{Mukherji19PRM}.
In this work, we have also validated that the PLA parameters give the correct estimates of specific volume 
at temperature of $T = 300$ K (i.e., $8.06 \times 10^{-4}~{\rm m^3/kg}$ for experiments and $8.11 \times 10^{-4}~{\rm m^3/kg}$ for simulations), 
elastic modulus, and thermal properties. The latter two will be discussed at a later stage in this draft.
The GROMACS molecular dynamics package \cite{gro} is used for the simulations.

The simulations in this work are performed in four steps with different chain lengths $N_{\ell}$ and number of chains $N_{\rm c}$.
In the following we give the details:

\begin{itemize}

	\item {\it Set 1, the KB analysis of blends}: For this step we have chosen $N_{\ell} = 3$, $N_{\rm c} = 1000$
		and six configurations within the range $0.0 \le x_{\rm LA} \le 1.0$. Each configuration is 
		equilibrated for 500 ns and the last 200 ns data is used for the calculation of ${\rm g}_{\rm ij}(r)$
		between different solution components. Here, ${\rm g}_{\rm ij}(r)$s are calculated between the
		center-of-masses of the central monomer of the individual trimers. The simulations are performed at a melt
		temperature of $T = 600$ K. 

	\item {\it Set 2, potential of mean force {\rm(}PMF{\rm )} using the umbrella sampling}: For these simulations, we have used monomeric
		fluid of MMA and LA at a 50-50 mixing ratio. A configuration consists of 6000 monomers and at $T = 600$ K.
		We have deliberately chosen a monomeric fluid to avoid any possible conformational fluctuation that might 
		lead to bias in the calculation of PMF. Here, we note in passing that PMF obtained from a monomeric system can be translated into 
		the monomers connected in a chain molecule using a treatment proposed by one of us with 
		collaborators \cite{mukherji17jcp}.

	\item {\it Set 3, morphology of blends}: In this case we have chosen $N_{\ell} = 30$, $N_{\rm c} = 200$
                and five configurations within the range $0.0 \le x_{\rm LA} \le 1.0$. Here, the 
		individual configurations are equilibrated for $1~\mu$s. The simulations are performed for the
		melt state at $T = 600$ K.

	\item {\it Set 4, mechanics and thermal transport calculations}: In this step, we take the 
		equilibrated configurations from the set 3 and then instantaneously quench these systems 
		to $T=300$ K where the stress-strain behavior and the thermal transport coefficients 
		$\kappa$ are calculated.
		
\end{itemize}

During $NpT$ simulations, the temperature is imposed using the velocity rescale 
thermostat with a coupling constant of 1 ps \cite{Vscale}.
The pressure is set to 1 atm with a Berendsen barostat using a time constant 0.5 ps~\cite{Berend}.
Electrostatics are treated using the particle-mesh Ewald method \cite{PME}. 
The interaction cutoff for the nonbonded interactions is chosen as $r_c = 1.0$ nm, which is of the same range as the typical 
correlation length in these systems \cite{mukherji19mac}. The simulation time step is chosen as $\Delta t = 1$ fs and the equations of 
motion are integrated using the leap-frog algorithm.

\subsection{Monomer-monomer potential of mean force}

To investigate the relative interaction strengths between different solution components, we have calculated PMF. 
Here, PMFs between two monomers is calculated using the umbrella sampling method \cite{valleau77} over a series of three independent 
simulations, i.e., between PMMA-PMMA, PLA-PLA and PMMA-PLA. For each simulation, a configuration is 
selected after 50 ns where a monomer type i is pulled apart from another monomer type j using the steered molecular dynamics,
with a spring constant $k = 1000~{\rm kJ \ mol^{-1} \ nm^{-2}}$ and at a pulling rate of $10^{-3}~{\rm nm \ ps^{-1}}$ over 1.5 ns.
From these simulations, we have selected twelve configurations within the range 0.35 nm $< r <$ 1.55 nm and with $\triangle r = 0.10$ nm.
The force distributions is obtained for each of the twelve distances over 5 ns long simulation each.
Subsequently, the weighted histogram analysis method (WHAM) was used to combine the force distributions into 
a potential of mean force (PMF) \cite{kumar92,spoel10}.

\subsection{Thermal transport calculations}

$\kappa$ is calculated using the approach-to-equilibrium method \cite{ATE}. 
For this purpose, we divide the simulation box into three compartments along the $x-$direction.
Here, the middle slab of width $L_x/2$ is sandwiched between two side slabs each having 
the width of $L_x/4$. Here, we employ periodic boundary conditions in all directions.

Simulations are performed in two stages: during the initial stage of canonical simulations 
the middle slab is kept at $T_{\rm Hot} = 350$ K, while the two 
side slabs are maintained at $T_{\rm Cold} = 250$ K. 
This simulation is performed for 5 ns each configuration. 
After this stage, the temperatures are allowed to relax during a microcanonical run 
for 50 ps. 

From the relaxation of $T_{\rm Hot}-T_{\rm Cold}$ during the microcanical simulations, 
we extract a bi-exponential fit and obtain the longitudinal time constants $\tau_x$. 
$\kappa$ can then be calculated using $\kappa = C_v L_x/4\pi^2L_yL_z\tau_x$. 
Here, we consider the Dulong-Petit classical estimate of the specific heat, i.e., $C_v = 3N_{\rm total}k_{\rm B}$, 
where $N_{\rm total}$ is the total number of atoms in the simulation box. 

\section{Results and discussions}
\label{sec:res}

\subsection{Radial distribution function and Kirkwood-Buff integral}

\begin{figure}[ptb]
\includegraphics[width=0.43\textwidth,angle=0]{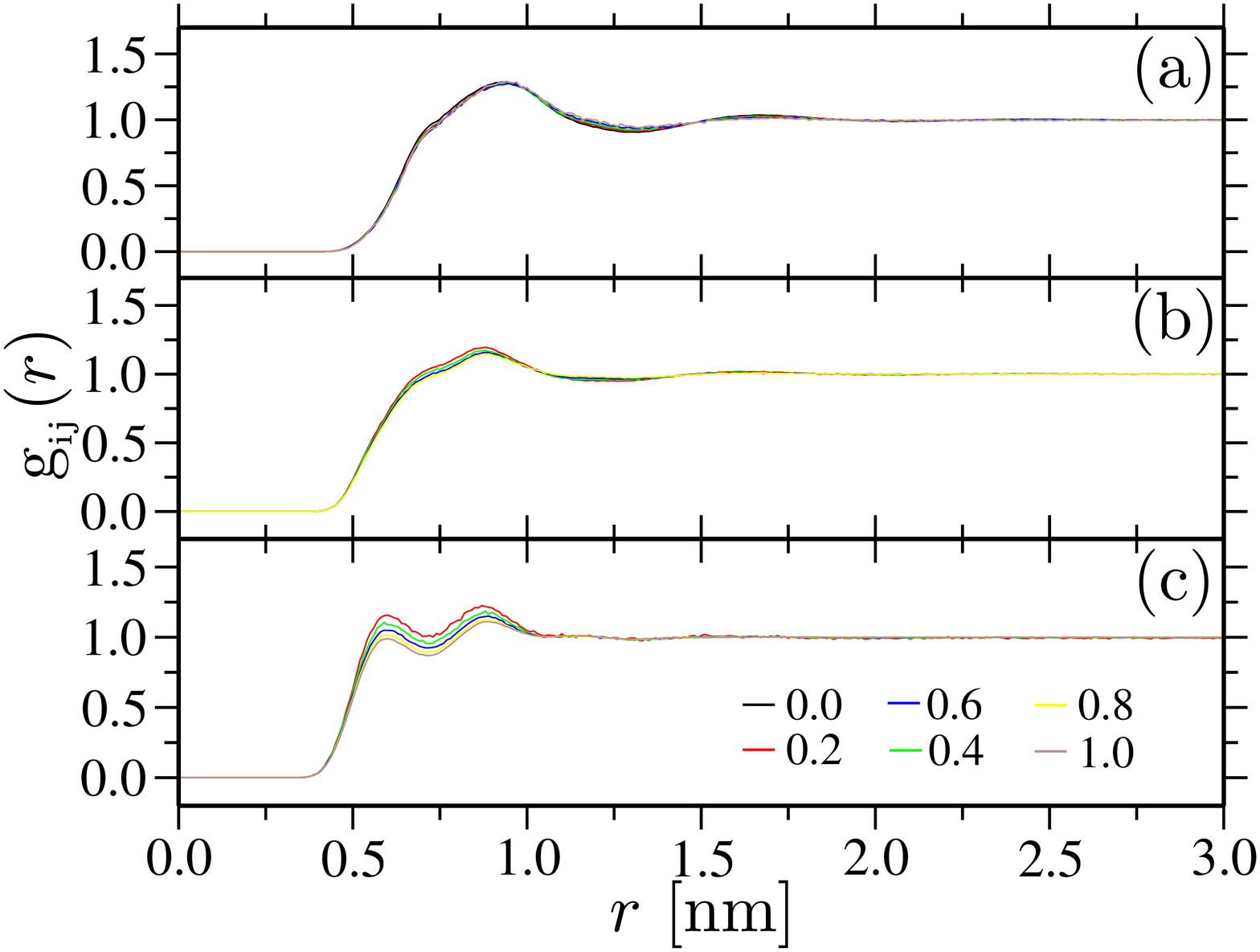}
\includegraphics[width=0.43\textwidth,angle=0]{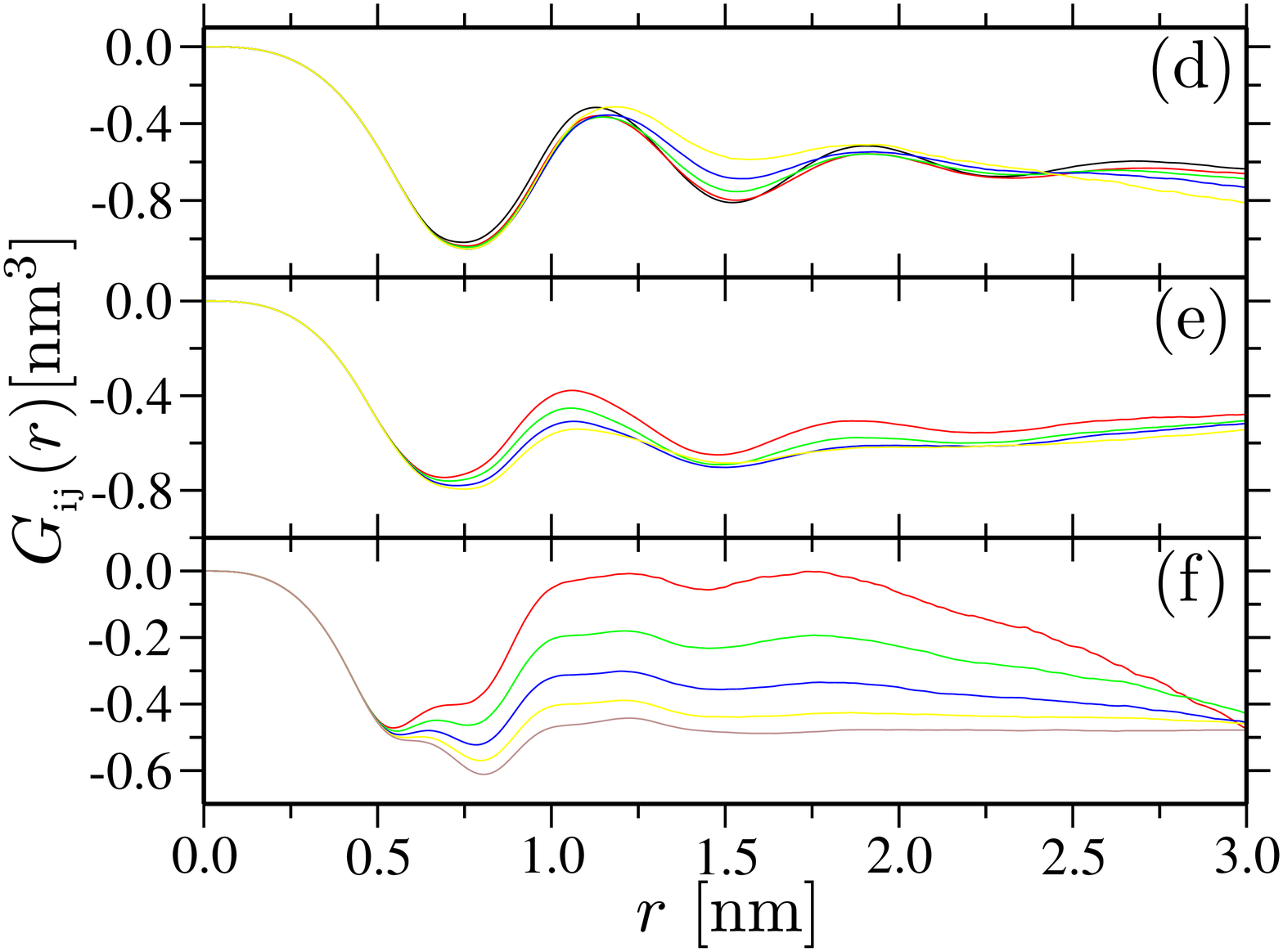}
	\caption{The pair distribution function ${\rm g}_{\rm ij}(r)$ (a-c) and the cumulative Kirkwood-Buff integral $G_{\rm ij}(r)$ (d-f)
	between different solution components of poly(methyl methacrylate) (PMMA) and poly(lactic acid) (PLA) blends for different 
	PLA mole fractions $0.0 \le x_{\rm LA} \le 1.0$. Data is shown from the simulations of trimers and ${\rm g}_{\rm ij}(r)$ and 
	$G_{\rm ij}(r)$ are calculated between the central monomers, namely MMA-MMA (a and d), MMA-LA (b and e) and LA-LA (c and f).
	%Panel (g) shows $G_{\rm ij}$ by taking the average of $G_{\rm ij}(r)$ within the interval $1.0{\rm nm} < r < 2.0{\rm nm}$.
	The temperature is set at $T = 600$ K and at 1 atm pressure.
	\label{fig:rdf}}
\end{figure}

In Fig.~\ref{fig:rdf} we show ${\rm g}_{\rm ij}(r)$ and ${G}_{\rm ij}(r)$ for different $x_{\rm LA}$. 
It can be appreciated that the LA$-$LA structures are mostly affected by the change in $x_{\rm LA}$, see Figs.~\ref{fig:rdf}(c) and ~\ref{fig:rdf}(f),
while the structures between the other pairs remain mostly unaffected within the error bar.
To obtain $G_{\rm ij}$ we average individual $G_{\rm ij}(r)$ within the range 1 nm $<r<$ 2 nm. 
The deviation from the plateau for $r>$ 2 nm is a typical behavior observed for $G_{\rm ij}(r)$ 
in the mid-sized closed boundary simulation setups \cite{mukherji13mac}. 
This is because the aggregation of one molecular species, in one region of the simulation domain,
leads to the depletion of the same species elsewhere.
In this context, ideally ${G}_{\rm ij}(r)$ should be calculated in the grand canonical 
ensemble \cite{kbi,kbibook,mukherji13mac}. However, since most simulations are performed 
in the mid-sized closed boundary domains, ${G}_{\rm ij}(r)$ often suffers from the severe system size effects. 
Here, we find that $N_{\rm c}=1000$ is a good estimate to obtain the converged ${G}_{\rm ij}(r)$, 
where the equilibrium box sizes for all $x_{\rm LA}$ values are over 8 nm. These box dimensions are over five times the 
typical correlation lengths in these systems, i.e., $r \simeq 1.5$ nm. 
We also note in passing that there are more elegant methods to
estimate ${G}_{\rm ij}$ within a semi-grand canonical simulation setup \cite{mukherji13mac} and from the particle 
fluctuations \cite{robin16jcp}, which is, however, beyond the scope of the present study.

Using the estimates of ${G}_{\rm ij}$ we obtain $\Delta {\mathcal G}_{\rm mix}$ by 
double integrating Eq.~\ref{eq:gibbs} using the boundary condition, i.e., $\Delta {\mathcal G}_{\rm mix} = 0$ at
$x_{\rm LA} = 0.0$ and 1.0.
We have also calculated $\Delta {H}_{\rm mix}$ per monomer from $NpT$ simulations using Eq.~\ref{eq:enthalpy}.
From these two estimates we obtain $T \Delta S_{\rm mix}$.
\begin{figure}[ptb]
\includegraphics[width=0.49\textwidth,angle=0]{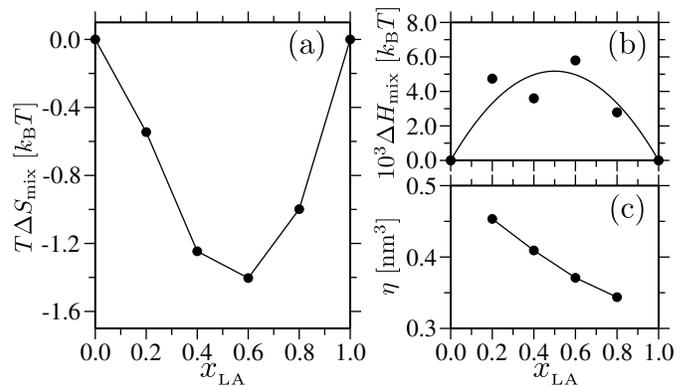}
	\caption{Parts (a) and (b) show the entropic $T \Delta S_{\rm mix}$ and the enthalpic $\Delta {H}_{\rm mix}$
	contributions to the change in the Gibbs free energy of mixing $\Delta {\mathcal G}_{\rm mix}$, respectively.
        Part (c) shows the preferential solvation parameter $\eta$. The data sets are shown for different 
	lactic acid (LA) mole fractions $x_{\rm LA}$. The lines in parts (a) and (c) are drawn to guide the eye,
	while the line in the part (b) is a fit to the data following $\chi x_{\rm LA}\left(1-x_{\rm LA}\right)$ with
	$\chi = 0.021k_{\rm B}T$.
\label{fig:femix}}
\end{figure}
In Figs.~\ref{fig:femix}(a-b) we show $T \Delta S_{\rm mix}$ and $\Delta {H}_{\rm mix}$ as a function of $x_{\rm LA}$.
It can be seen that the entropic contribution (that favors mixing) dominates over the enthalpy (that favors
demixing because it is positive). This indicate that the mixing of PMMA-PLA systems is 
entropically stabilized. This is also consistent with the recent experimental results \cite{hou21maclet}. 
From the data in Fig.~\ref{fig:femix}(b), we also obtain $\chi-$parameter 
using the FH function $\Delta {H}_{\rm mix}=\chi x_{\rm LA}\left(1-x_{\rm LA}\right)$.
Here, we find $\chi = 0.021~k_{\rm B}T$ at the monomer level, which might look as a small value.
However, for large $N_{\ell}$ this can lead to a significantly large $N_{\ell}\chi$.
Additionally, $T \Delta S_{\rm mix} \propto 1/N_{\ell}$, see Eq.~\ref{eq:entropy}. 
Therefore, within the long chain limit, one is expected to observe a clear demixing in 
these systems, as predicted by the FH theory \cite{doibook,genbook}. However,
for the moderate $N_{\ell}$ values, weak miscibility is expected. We will come back to this 
point in the next subsection.

A fit to the data in Fig.~\ref{fig:femix}(b) gives $\chi$. This is a good estimate so long as the mixing 
of two species does not introduce any nonlinear variation in the total number density $\rho$. 
However, in most chemical systems, as in the case of PMMA-PLA blends, $\rho$ values usually show 
a weak non-linear variation with $x_{\rm LA}$ (data not shown). Therefore, a more reasonable picture will be to
have a concentration dependent $\chi-$parameter. In this context, a quantity that can give an estimate 
of $x_{\rm LA}$ dependent effective relative interaction strengths is the preferential solvation parameter $\eta$.
In Fig.~\ref{fig:femix}(c) we shown $\eta$ as a function of $x_{\rm LA}$, which also indicates that $\eta > 0$ for all $x_{\rm LA}$.

\begin{figure}[ptb]
\includegraphics[width=0.43\textwidth,angle=0]{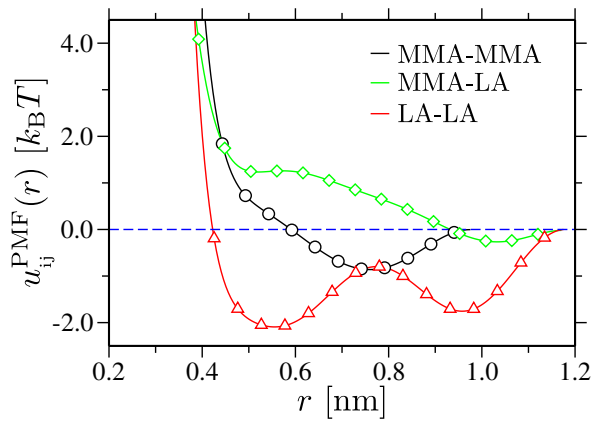}
	\caption{Potential of mean force $u^{\rm PMF}_{\rm ij}(r)$ between different solution components of methyl methacrylate (MMA)
	and lactic acid (LA) at a LA mole fractions $x_{\rm LA} = 0.5$ and at a temperature $T = 600$ K. Symbols are only shown for every 10th 
	data point. The data is shown from the simulations of the monomeric fluids, where $u^{\rm PMF}_{\rm ij}(r)$ 
	can be translated from an isolated monomer system to a monomer in a chain
	using ${\tilde u}_{\rm ij}^{\rm PMF}(r) =  - k_{\rm B}T [\ln \{e^{-u_{ij}^{\rm PMF}(r)/k_{\rm B}T} + 1\} - \ln \left(2\right)]$.
\label{fig:pmf}}
\end{figure}

The data in Figs.~\ref{fig:femix}(b-c) also indicate that the interactions between the like species are preferred
over the cross-interactions. To quantify these relative interaction strengths, we have also calculated the monomer-monomer
PMFs between different solution species $u^{\rm PMF}_{\rm ij}(r)$, see Fig.~\ref{fig:pmf}. It can be appreciated that the interactions between LA$-$LA
and MMA$-$MMA is preferred over the cross-interactions. It is important to highlight that we have only calculated $u^{\rm PMF}_{\rm ij}(r)$
for $x_{\rm LA} = 0.5$ because pair structures are mostly unaffected by $x_{\rm LA}$, except for ${\rm g}_{\rm LA-LA}(r)$, 
and thus $u^{\rm PMF}_{\rm ij}(r)$ is a good estimate at $x_{\rm LA} = 0.5$.

\subsection{Morphology}

%{\bf Make connections with miscibility from trimers and polymers ..... Fig.5}

Experiments have shown that the PMMA-PLA blends form 
``so called'' co-continuous phases \cite{le06,hou21maclet}. The typical length scales in these phases were 
found to be of the order of $1-2~\mu$m \cite{hou21maclet}. It has been argued that$-$ while $\Delta H_{\rm mix}$ favors demixing, 
the morphologies are entropically stable. This trend is consistent with our simulation data presented 
in Fig.~\ref{fig:femix}. Therefore, it will also be important to study how the observed trends 
in Fig.~\ref{fig:femix} can be used to understand the morphologies of PMMA-PLA mixtures in simulations. Here, however, it is important to emphasize that by no means
we aim to claim anything close to the experimental length scales (both in terms of the system dimensions and
the chain lengths) due to the obvious limitations associated with the molecular simulations. The goal of this work is rather to see if the 
simulations can show some qualitative features observed in the experiments.

For these simulations, we have chosen $N_{\ell}=30$ for both PMMA and PLA. Here, PMMA 
has a persistence length $\ell_{\rm p} \simeq 0.65$ nm and PLA has $\ell_{\rm p} \simeq 0.50$ nm.
This will give $N_{\ell}=12\ell_{\rm p}$ for PMMA and $N_{\ell}=15\ell_{\rm p}$ for PLA.
While $N_{\ell}=12-15\ell_{\rm p}$ might look as too short to study the polymer properties, 
it is known to reasonably capture the conformational behavior of a single chain in solution \cite{Mukherji17NC} and 
in melt \cite{Mukherji19PRM}. 

\begin{figure}[ptb]
\includegraphics[width=0.49\textwidth,angle=0]{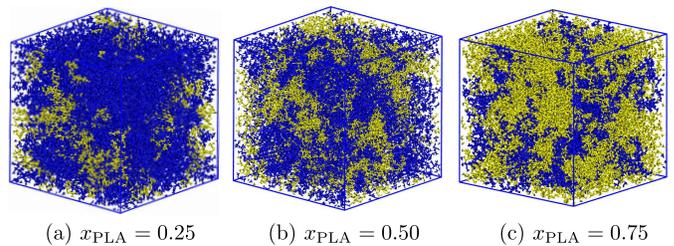}
	\caption{Simulation snapshots of the poly(methyl methacrylate) (PMMA) and poly(lactic acid) (PLA) 
	blends for three different PLA mole fractions $x_{\rm LA}$. The data is shown for a chain length
	$N_{\ell}=30$ and for the melt state at a temperature of $T = 600$ K. PMMA and PLA chains are
	rendered in blue and yellow, respectively. The system dimensions $L$ for $x_{\rm LA} = 0.25$,
	0.50, and 0.75 are 9.76 nm, 9.45 nm and 9.12 nm, respectively. 
\label{fig:snapequil}}
\end{figure}

In Fig.~\ref{fig:snapequil} we show simulation snapshots for three different $x_{\rm PLA}$.
Configurations are shown after $1~\mu$s equilibration for each system. We would also like to
highlight that the typical end-to-end relaxation time of a chain in a PMMA-PLA melt is only 
about 20$-$25 ns for $N_{\ell} = 30$ and thus the individual chains have moved 40$-$50 times its own size 
$R_{\rm g} \simeq 1.5$ nm over $1~\mu$s.
From the simulations, it is evident that the systems do not phase separate with the 
well-defined PLA and PMMA domains, rather form phases with PLA forming percolating network-like 
structures within the blends. While it is quite evident that the study of phase separation 
within an all-atom setup is nontrivial task especially for the long chain molecules, 
our system dimensions are only about $6R_{\rm g}$ and thus the systems show weak miscibility 
that is entropically stabilized. Another quantity that speaks in the favor that the systems do not
phase separate (as predicted by Fig.~\ref{fig:femix} and Fig.~\ref{fig:pmf}) 
is that $G_{\rm ij}(r)$ (in Figs.~\ref{fig:rdf}(d-f)) show convergence. For the phase separated 
systems, $G_{\rm ij}(r)$ should diverge for the large $r$ values.

We also note in passing that a completely immiscible polymer blend, especially consisting of the 
brittle polymers, would have been extremely unstable under any external deformation.
Therefore, to further consolidate our argument of PMMA-PLA phase behavior with percolating PLA chains, 
we will now investigate the mechanical response of the configurations presented in Fig.~\ref{fig:snapequil}. 

\subsection{Mechanical response and thermal transport}

\begin{figure}[ptb]
\includegraphics[width=0.49\textwidth,angle=0]{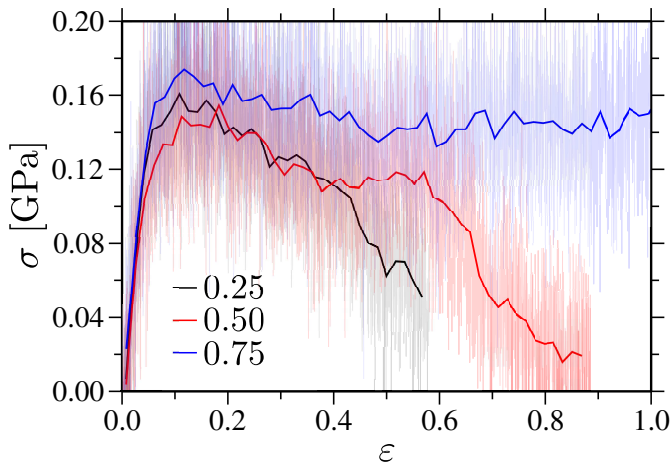}
	\caption{Stress $\sigma$ as a function of strain $\varepsilon$ of the poly(methyl methacrylate) (PMMA) and poly(lactic acid) (PLA)
	blends for three different PLA mole fractions $x_{\rm PLA}$. The simulation box is subjected to an uniaxial tensile deformation 
	along the $z-$axis with a velocity of $10^{-3}$ nm/ps. The background shaded data sets are from the raw data and
	for the representation purpose we also draw the block averaged data (solid lines). The data is shown 
	at a temperature $T=300$ K.
\label{fig:ss}}
\end{figure}

\begin{figure}[ptb]
\includegraphics[width=0.49\textwidth,angle=0]{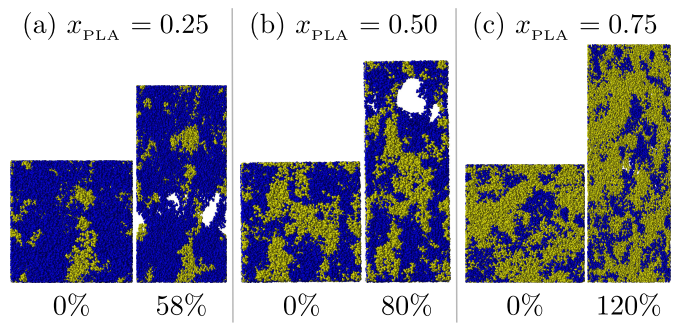}
	\caption{Simulation snapshots showing the undeformed and at the onset of fracture configurations for three
	different PLA mole fractions $x_{\rm PLA}$. The data is shown 
        at a temperature $T=300$ K.
\label{fig:fracture}}
\end{figure}

In Fig.~\ref{fig:ss} we show the stress-strain behavior for three different $x_{\rm PLA}$ in their
glassy state at $T=300$ K.
It can be appreciated that the individual data sets clearly show the typical behavior: 
namely a linear response for $\varepsilon<3\%$, a yield peak at around $\varepsilon \simeq 15\%$, 
a plastic deformation for $\varepsilon > 20\%$, and an ultimate failure with $x_{\rm PLA}-$dependent strain-to-break $\varepsilon_{\rm f}$, which is defined as strain where the $\sigma$ values depart from the plateau in Fig.~\ref{fig:ss}. Here, $\varepsilon_{\rm f} = 0.34$ for $x_{\rm PLA}=0.25$, $\varepsilon_{\rm f} =  0.55$ for $x_{\rm PLA}=0.50$ and $\varepsilon_{\rm f} = 1.38$ for $x_{\rm PLA}=0.75$ \cite{statement}.
Fig.~\ref{fig:ss} also shows that the systems become more ductile with increasing $x_{\rm PLA}$, attaining 
the largest ductility for $x_{\rm LA}=0.75$ with $\varepsilon_{\rm f} > 100\%$. 
This is also evident from the simulation snapshots in Fig.~\ref{fig:fracture}. 
The observation in Fig.~\ref{fig:ss} is in qualitative agreement with the experimentally observed trends~\cite{hou21maclet}.

What causes this increasing ductility with the increase in $x_{\rm PLA}$? To discuss this point, we will now put
forward different scenarios:

It is a common practice that a brittle polymeric material (such as PMMA) is usually plasticized by blending 
in another polymer with a lower glass transition temperature $T_{\rm g}$ \cite{thesis,eranrev}. 
In that case, the enhancement of ductility would then be due to the possible enhancement of mobility in the vicinity of $T_{\rm g}$.
Here, a pure PLA material has a lower $T_{\rm g}$ than a pure PMMA sample~\cite{phbook}. Furthermore, $T_{\rm g}$ changes monotonically with $x_{\rm PLA}$ \cite{gar18,hou21maclet}, 
i.e., the larger the $x_{\rm PLA}$, the lower the $T_{\rm g}$. Within this argument, if
$T$ of a glassy state is chosen such that $T_{\rm g}-T$ is small (as in the case of the PMMA-PLA blends,
i.e., 35 K $< \left(T_{\rm g}-T\right) < $ 75 K) and thus the elastic response gets highly dependent on 
$x_{\rm PLA}$ (or $T_{\rm g}-T$).
Within this picture, if $T_{\rm g}-T$ is increased, a system might become less ductile and can even show $\varepsilon_{\rm f}$ closer to that of as observed for the smaller $x_{\rm PLA}$. To test this hypothesis, we have performed one set of simulations 
with $T = 200$ K and $x_{\rm PLA} = 0.75$. In this case $\varepsilon_{\rm f}$ changes only about 5\% with respect to $T=300$ K.
Here, it is important to highlight that a system deep in the glassy state, as in the case for $T = 200$ K,
there still exists a contrast in the elastic modulus between the two pure phases,
which is, however, only about 15-20\% between a pure PMMA and a pure PLA systems and thus is not so significant to make a large difference in the ductility. These observations speak against any explanation of increased ductility purely on $T_{\rm g}$.

Another possible scenario is based on the details of chemical structures of individual monomer units, see Fig.~\ref{fig:schem}. In this
case, the side groups of PMMA can strongly interact and thus increase inter-chain interlocking, thus makes PMMA a brittle system. On the contrary, from a purely structural point of view, a PLA chain can slide past another PLA or a PMMA chain with rather small friction. Therefore, increasing $x_{\rm PLA}$ can withstand larger $\varepsilon$ upon deformation, while the chains percolate to form the elongated network-like structures. This is seen by comparing two snapshots in Fig.~\ref{fig:fracture}(c). This suggests a delicate interplay between the monomer structures, the local packing and the mechanical response.

Here it is also important to highlight that upon deformation small free spaces appear that coalesce to fracture the samples \cite{mukherji08pre}. This coalescence happens at a larger $\varepsilon$ with increasing $x_{\rm PLA}$ values. Furthermore, PLA chains can also join hands (or intermingle) because of their preferred monomer-monomer interaction, see Fig.~\ref{fig:pmf}. Therefore, a PLA chain can move more easily with respect to a PMMA chain and thus can rearrange itself more easily when put under a tensile deformation. To investigate this aspect, we have also performed two additional simulations with shorter PLA chains $N_{\rm PLA} = 10$ and 20, while keeping the chain length of PMMA fixed, i.e., $N_{\rm PMMA} = 30$. Note that our chain lengths are short and therefore we are deliberately avoiding to use ``entanglement" in our nomenclature above (instead used intermingle), which is a well defined concept in the long chain polymer melts.

\begin{figure}[ptb]
\includegraphics[width=0.49\textwidth,angle=0]{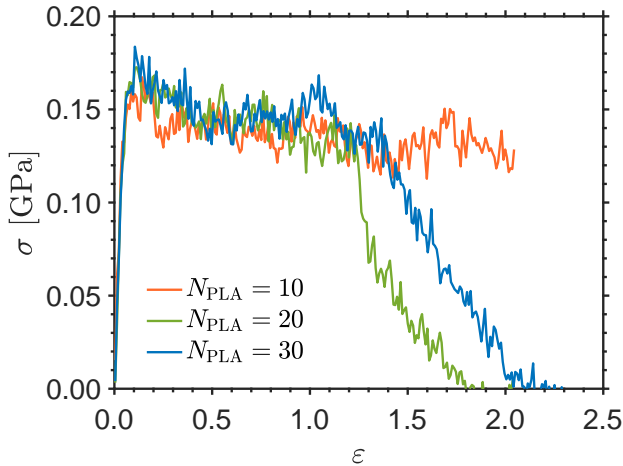}
	\caption{Same as Fig.~\ref{fig:ss}, however, for one poly(lactic acid) (PLA) mole fraction $x_{\rm PLA} = 0.75$ and for three different PLA chain lengths $N_{\rm PLA}$. %For better representation, we have only plotted the data for $1.0 < \varepsilon < 2.5$.
\label{fig:ss2}}
\end{figure}

In Fig.~\ref{fig:ss2}, we show the typical stress-strain behavior for three different $N_{\rm PLA}$ and for $x_{\rm PLA} = 0.75$. 
It can be appreciated that $\varepsilon_{\rm f}$ shows a non-monotonic dependence with $N_{\rm PLA}$, i.e., going from $N_{\rm PLA} = 30$ to 20, $\varepsilon_{\rm f}$ first decreases by about 10-15\% and then for $N_{\rm PLA} = 10$, $\varepsilon_{\rm f}$ increases again. A simple picture would then be$-$ increasing $N_{\rm PLA}$ mean that a chain will read a fully stretched conformation at a larger $\varepsilon_{\rm f}$. This, however, contradicts the increased ductility at $N_{\rm PLA} = 10$. Here it noteworthy that in the case of perfect miscibility this non-monotonic ductility would be quite obvious because of the chain bridging/gluing the broken interfaces and chain rearrangements, a similar scenario as observed in the thermoset-thermoplastic blends \cite{mukherji09epl}. On the contrary, however, PLA-PMMA blends are partially miscible as expected from the solvation thermodynamics, see Fig.~\ref{fig:femix}. In this context, we speculate that
%while the decrease in $\varepsilon_{\rm f}$ between $N_{\rm PLA} = 20$ and 30 is may just be within the error bar, 
the significant increase in $\varepsilon_{\rm f}$ for $N_{\rm PLA} = 10$ (which is only about 3 persistence lengths $\ell_{\rm p}$) might be due to the fact that the shorter chains act like single particles and thus can rearrange faster, thus hindering the coalescence of the free volumes within the samples. 
%It is also noteworthy that $N_{\rm PLA} = 10$ is only about 3 persistence lengths $\ell_{\rm p}$ and thus can hardy have any conformational fluctuation.
These results highlight the importance of chain friction in the brittle-to-ductile transition of PLA-PMMA blends.
%Given the observation in Fig.~\ref{fig:ss2}, we can not state at what exact $N_{\rm PLA}$ will $\varepsilon_{\rm f}$ increase. However, w
We also note in passing that$-$ the numerical validation of the above scenarios will require a set of careful investigations with a variety of monomer structures, monomer-monomer interactions, polymer blends, chain lengths, and mixing ratios. In particular, to investigate how the delicate balance between the molecular flexibility, the monomer structure, the detailed interactions and the blend morphologies can dictate the mechanical response of the commodity blends, which will, however, be presented elsewhere \cite{mukherjiPrep}.

\begin{figure}[ptb]
\includegraphics[width=0.49\textwidth,angle=0]{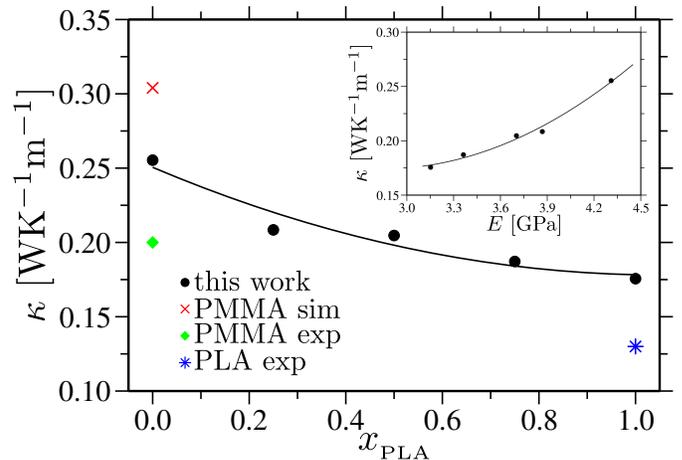}
	\caption{The main panel shows the thermal transport coefficient $\kappa$ as a function of poly(lactic acid) (PLA)
	mole fraction $x_{\rm PLA}$. For comparison, we have also included experimental data for PMMA \cite{cahill16mac} 
	and PLA \cite{phbook} and also simulation data for PMMA from an earlier work using the M\"uller-Plathe method \cite{Mukherji19PRM}. 
	Inset shows the variation of $\kappa$ with the elastic modulus $E$. Lines are drawn to guide the eye.
	The data is shown for a temperature $T=300$ K.
\label{fig:kappa}}
\end{figure}

Another quantity that can be extracted from Fig.~\ref{fig:ss} is the elastic modulus $E$ from the linear regime, i.e., for $\varepsilon<3\%$. 
In our simulations, going from pure PLA to PMMA, we find 3.0 GPa $< E <$ 4.3 GPa. These values are also within the similar 
range as observed in the experiments \cite{phbook,cahill16mac}. A quantity that is directly related to $E$ 
is the thermal transport coefficient $\kappa$ of the amorphous polymers, as predicted by the minimum thermal
conductivity model $\kappa \propto E$ \cite{cahill92prb}. Therefore, we have also calculated $\kappa$ 
and the corresponding data is shown in the main panel of Fig.~\ref{fig:kappa} and the 
dependence of $\kappa$ on $E$ is shown in the inset of Fig.~\ref{fig:kappa}.
As expected from the blending of two materials, $\kappa$ with $x_{\rm PLA}$ shows a monotonic variation between
the $\kappa$ values of a pure PLA and a pure PMMA systems. While $\kappa$ values are known for
the pure phases on PMMA and PLA, which is also consistent with the earlier experiments \cite{phbook,cahill16mac} 
and simulations \cite{Mukherji19PRM}, we report $\kappa$ data for PMMA-PLA blends here.
%. Furthermore, as predicted by the minimum thermal
%conductivity model \cite{cahill92prb}, $\kappa \propto E$ (see the inset in Fig.~\ref{fig:kappa}).
This shows an internal correlation between the linear elasticity, mechanics, and thermal transport of 
commodity polymer blends. Here, we also note in passing that the earlier studies have reported that the
symmetric blends of the long chain macrmolecules can show the monotonic variation in $\kappa$ between
two pure phases \cite{cahill16mac}, while a weak non-monotonic variation in $\kappa$ with relative 
concentration was observed for the asymmetric blends \cite{mukherji19mac}. Here, PMMA-PLA blends fall 
under the former catagory.

\section{Conclusions}
\label{sec:conc}

Using large scale molecular dynamics simulation of an all-atom model, we have established an internal 
correlation between the solvation thermodynamics, the blend morphology, the non-linear mechanics and the thermal transport 
in a commodity polymer blend. For this purpose, we have chosen a technologically relevant blends of
the commodity polymers, namely poly(methyl methacrylate) (PMMA) and poly(lactic acid) (PLA).
Studying the thermodynamics in polymer mixtures is rather nontrivial task. Therefore, we make use of the fluctuation theory of 
Kirkwood and Buff \cite{kbi} that connects the pair-wise structure to the mixture thermodynamics. We highlight
the importance of the delicate balance between the entropic $T\Delta S_{\rm mix}$ and the enthalpic $\Delta H_{\rm mix}$ 
contributions to the mixing Gibbs free energy change $\Delta {\mathcal G}_{\rm mix}$. 
We further show how these contributions can be individually extracted within the molecular dynamics simulations,
especially for the long chain macromolecular mixtures. These observations are correlated with the morphologies of 
PLA-PMMA blends and finally to the mechanical and thermal responses. 

Even when we have only presented results for one system, our work achieves three key highlights: 
(1) Establish a structure-property relationship within one unified simulation framework, (2) a possible 
explanation of the unexpected mechanical response of the PMMA-PLA blends, as observed in a
recent experiment \cite{hou21maclet}, and (3) provide a set of possible scenarios that can serve as 
a guiding tool to understand polymer blends. Moreover, further works need to perform on a broad range of polymer blends with 
different microscopic interactions, miscibility, monomeric structures and chain conformations.
However, the hand-in-hand correlation presented in this work between different physical properties is rather rare and the
protocol presented may be used to design the multi-component materials 
for their advanced functional applications. 

\section{Acknowledgement}
D.M. further thanks the ARC Sockeye facility of 
the University of British Columbia, the Compute Canada facility and 
the Quantum Matter Institute LISA cluster, where major part of the simulations were performed. 
This research was undertaken thanks, in part, to the Canada First Research Excellence Fund (CFREF), 
Quantum Materials and Future Technologies Program.
T.E.O. thanks the Centro Nacional de Supercomput\c{c}\~ao (CESUP) for providing HPC resources, where part of the simulations were performed.
C.R. thanks MITACS for the financial support.


\begin{thebibliography}{55}

\bibitem{Kroeger04PR}
M. K\"oger,
{\it Physics Reports} {\bf 390}, 453 (2004).

\bibitem{Mueller20PPS}
M. M\"uller, 
{\it Progress in Polymer Science} {\bf 101}, 101198 (2020).

\bibitem{Mukherji20AR}
D. Mukherji, C. M. Marques, and K. Kremer, 
{\it Annual Reviews of Condensed Matter Physics} {\bf 11}, 271 (2020).		

\bibitem{stevens01mac}
M. J. Stevens,
{\it Macromolecules} {\bf 34}, 2710 (2001).

\bibitem{mukherji08pre}
D. Mukherji and C. F. Abrams,
{\it Phys. Rev. E} {\bf 78}, 050801 (2008).

\bibitem{white10anrev}
B. J. Blaiszik, S. L. B. Kramer, S. C. Olugebefola, J. S. Moore, N. R. Sottos, S. R. White,
{\it Annual review of materials research} {\bf 40}, 179 (2010).

\bibitem{palmese14jmat}
M. Sharifi, C. W. Jang, C. F. Abrams and G. R. Palmese,
{\it J. Mater. Chem. A} {\bf 2}, 16071 (2014).

\bibitem{pipe14nm}
G. Kim, D. Lee, A. Shanker, L. Shao, M. S. Kwon, D. Gidley, J. Kim, and K. P. Pipe,
{\it Nat. Mater.} {\bf 14}, 295 (2015).

\bibitem{xie16mac}
X. Xie, D. Li, T. Tsai, J. Liu, P. V. Braun, and D. G. Cahill,
{\it Macromolecules} {\bf 49}, 972 (2016).

\bibitem{mukherji19mac}
D. Bruns, T. E. de Oliveira, J. Rottler, and D. Mukherji,
%Tuning morphology and thermal transport of asymmetric smart polymer blends by macromolecular engineering.
{\it Macromolecules} {\bf 52}, 5510 (2019).

\bibitem{kim05apl}
N. Kim, B. Domercq, S. Yoo, A. Christensen, B. Kippelen, and S. Graham, 
{\it Appl. Phys. Lett.} {\rm 87}, 241908 (2005).

\bibitem{cola16aami}
M. K. Smith, V. Singh, K. Kalaitzidou, and B. A. Cola,
{\it ACS Appl. Mater. Int.} {\bf 8}, 14788 (2016).

\bibitem{hou21maclet}
X. Hou, S. Chen, J. J. Koh, J. Kong, Y.-W. Zhang, J. C. C. Yeo, H. Chen, and C. He
		{\it ACS Macro Lett.} {\bf 10}, 406 (2021).

	\bibitem{yeo20}
J. C. C. Yeo, J. K. Muiruri, J. J. Koh, W. Thitsartarn, X. Zhang, J. Kong, T. T. Lin, Z. Li, C. He, 
%Bend, Twist, and Turn: First
%Bendable and Malleable Toughened PLA Green Composites. 
{\it Adv. Funct. Mater.} {\bf 30}, 2001565 (2020).

	\bibitem{paul03}		
P. P\"otschke and D. R. Paul,
		{\it J. Macromol. Sci., Polym. Rev.} {\bf 43}, 87 (2003).

\bibitem{le06}
K.-P. Le, R. Lehman, J. Remmert, K. Vanness, P. M. L. Ward, and J. Idol,
		{\it J. Bio. Sci. Pol. Ed.} {\bf 17}, 121 (2006).
	
	\bibitem{sam15}
C. Samuel, J.-M. Raquez, and P. Dubois,
		{\it AIP Conf. Pro.} {\bf 1664}, 030005 (2015).

	\bibitem{gar18}
		M. Gonzalez-Garzon, S. Shahbikian, and M. A. Huneault,
		{\it J. Pol. Res.} {\bf 25}, 58 (2018).

\bibitem{doibook}
M. Doi,
{\it Soft Matter Physics} Oxford University Press (2013).

\bibitem{genbook}
P.-G. de Gennes,
{\it Scaling Concepts in Polymer Physics} Cornell University Press, London (1979).

\bibitem{kbi}
J. G. Kirkwood and F. P. Buff,
{\it J Chem. Phys.} {\bf 19}, 774 (1951).

\bibitem{kbibook}
A. Ben Naim,
{\it Molecular Theory of Solutions} Oxford University Press, London (2006).

\bibitem{robin16jcp}
R. Cortes-Huerto, K. Kremer, and R. Potestio,
{\it J Chem. Phys.} {\bf 145}, 141103 (2006).

\bibitem{doros18fluid}
A. A. Galata, S. D. Anogiannakis, and D. N. Theodorou,
{\it Fluid Phas. Equil.} {\bf 470}, 25 (2018).

\bibitem{robin21arx}
M. Sevilla and R. Cortes-Huerto,
{\it arXiv:2110.14536} (2021). 

\bibitem{smith07}
M. Kang and P. E. Smith,
{\it Fluid Phas. Equil.} {\bf 256}, 14 (2007).

\bibitem{pier08}
V. Pierce, M. Kang, M. Aburi, S. Weerasinghe, and P. E. Smith,
{\it Cell Biochem. Biophys.} {\bf 50}, 1 (2008).

\bibitem{mukherji13mac}
D. Mukherji and K. Kremer,
{\it Macromolecules} {\bf 46}, 9158 (2013).

\bibitem{kumari21pccp}
P. Kumari, V. V. S. Pillai, D. Gobbo, P. Ballone, and A. Benedetto,
{\it Phys. Chem. Chem. Phys.} {\bf 23}, 944 (2021).

\bibitem{OPLS}
        W. L. Jorgensen, D. S. Maxwell, and J. Tirado-Rives,
{\it J Am. Chem. Soc.} {\bf 118}, 11225 (1996).

\bibitem{Mukherji17NC}
        D. Mukherji, C. M. Marques, T. St\"uhn, and K. Kremer,
{\it Nat. Commun.} {\bf 8} {\bf 1374} (2017).

%\bibitem{Mukherji17JCP}
%T. E. de Oliveira, D. Mukherji, K. Kremer, and P. A. Netz.
%Journal of Chemical Physics 146:034904, 2017.

\bibitem{Mukherji19PRM}
C. Ruscher, J. Rottler, C. E. Boott, M. J. MacLachlan, and D. Mukherji,
{\it Phys. Rev. Mat.} {\bf 3}, 125604 (2019).

\bibitem{gro}
        S. Pronk, S. Pall, R. Schulz, P. Larsson, P. Bjelkmar, R. Apostolov, M. R. Shirts, J. C. Smith, P. M. Kasson, D. van der Spoel, B. Hess, and E. Lindahl,
	{\it Bioinformatics} {\bf 29}, 845 (2013).

\bibitem{mukherji17jcp}
C. De Silva, P. Leophairatana, T. Ohkuma, J. T. Koberstein, K. Kremer, and D. Mukherji,
{\it J Chem. Phys.} {\bf 147}, 064904 (2017).

\bibitem{Vscale}
        G. Bussi, D. Donadio, and M. Parrinello.
Canonical sampling through velocity rescaling.
{\it J Chem. Phys.} {\bf 126}, 014101 (2007).

\bibitem{Berend}
	H. J. C. Berendsen, J. P. M. Postma, W. F. van Gunsteren, A. DiNola, and J. R. Haak,
	{\it J Chem. Phys.} {\bf 81}, 3684 (1984).

\bibitem{PME}
        U. Essmann, L. Perera, M. L. Berkowitz, T. Darden, H. Lee, and L. G. Pedersen.
A smooth particle mesh ewald method.
{\it J Chem. Phys.} {\bf 103}, 8577 (1995).

\bibitem{valleau77}
G. M. Torrie, J. P. Valleau
{\it J. Comput. Phy.} {\bf 23}, 187 (1977).

\bibitem{kumar92}
S. Kumar, D. Bouzida, R. H. Swendsen, P. A. Kollman, J. M. Rosenberg
{\it J. Comput. Chem.} {\bf 13}, 1011 (1992).

\bibitem{spoel10}
J. S. Hub, B. L. de Groot, D. van der Spoel
{\it J. Chem. Theory Comput.} {\bf 6}, 3713 (2010).

\bibitem{ATE}
E. Lampin, P. L. Palla, P. A. Francioso, and F. Cleri, 
Thermal conductivity from approach-to-equilibrium molecular dynamics. 
{\it J. Appl. Phys.} {\bf 114} 033525 (2013).

\bibitem{statement}
Since the simulations domains are rather small (only about 8$-$9 nm), we have also performed simulations when the systems are elongated along the $y-$ and the $z-$directions to investigate if the deformation along only one direction introduces any morphological bias on the mechanical response. The mechanical responses along all three different directions only varry by less than 2\%.

\bibitem{thesis}
A. Bouzouita,
{\it Elaboration of polylactide-based materials for automotive application : study
of structure-process-properties interactions} PhD thesis (2016).

\bibitem{eranrev}
R. Long, C.-Y. Hui, J. P. Gong, and E. Bouchbinder,
{\it Annual Reviews of Condensed Matter Physics} {\bf 12}, 71 (2021).

\bibitem{phbook}
J. Brandrup, E. H. Immergut, and E. A. Grulke, 
{\it Polymer Handbook, 1079 2 Volumes Set, 4th ed.} Wiley (2003).

\bibitem{mukherji09epl}
D. Mukherji and C. F. Abrams,
{\it Euro Phys. Lett.} {\bf 88}, 56001 (2009).

\bibitem{mukherjiPrep}
D. Mukherji et al,
{\it in preparation} (2021).

\bibitem{cahill16mac}
X. Xie, D. Li, T. Tsai, J. Liu, P. V. Braun, and D. G. Cahill,
{\it Macromolecules} {\bf 49}, 972 (2016).

\bibitem{cahill92prb}
D. G. Cahill, S. K. Watson, and R. O. Pohl,
{\it Phys. Rev. B} {\bf 46}, 6131 (1992).

\end{thebibliography}
\end{document}